\documentclass[10pt]{article}
   \usepackage{epsfig}
   
\begin{document}

   \title{\LARGE \bf  Classification of String-like
   Solutions \\ in Dilaton Gravity\\}

   \author{
  \large  Y. Verbin$^a$\thanks{Electronic address:
verbin@oumail.openu.ac.il}$\:$,
  S. Madsen$^b$\thanks{Electronic address: madsen@imada.sdu.dk} $\:$,
 A.L. Larsen$^b$\thanks{Electronic address: all@fysik.ou.dk}$\:$
 and  M. Christensen$^b$\thanks{Electronic address: inane@fysik.sdu.dk}$\:$}
\date{ }
   \maketitle
 \centerline{$^a$ \em Department of Natural Sciences, The Open University
   of Israel,}
   \centerline{\em P.O.B. 39328, Tel Aviv 61392, Israel}
     \vskip 0.4cm
   \centerline{$^b$ \em Department of Physics, University of Odense, }
   \centerline{\em Campusvej 55, 5230 Odense M, Denmark}
   \vskip 1.1cm

   \begin{abstract}
   The static string-like solutions of the  Abelian Higgs model coupled 
  to dilaton gravity are analyzed and compared to the non-dilatonic case. 
  Except for a special coupling between the Higgs Lagrangian and  the  
dilaton, the solutions are flux tubes that generate a non-asymptotically 
 flat geometry.
   Any point in parameter space corresponds to two branches of solutions
with two
   different asymptotic behaviors. Unlike the non-dilatonic case, where
  one branch is always asymptotically conic, in the present case the
asymptotic
  behavior changes continuously along each branch. \\
   \end{abstract}


   \section{Introduction}
   \setcounter{equation}{0}
   \label{secintro}

Of all the topological defects \cite{vilsh}, which may have been formed during
phase transitions in the early universe, cosmic strings \cite{Vil0,KibbleH} are
those which have attracted most attention from a cosmological point of view. 
They were introduced into cosmology some 20 years ago by  Kibble 
\cite{Kibble1},  Zel'dovich \cite{Zel} and Vilenkin \cite{Vil1}, and were
considered for a long time as possible sources for density perturbations and
hence for structure formation in the universe. Indeed, the latest data from the
BOOMERANG and MAXIMA experiments \cite{BOOM,MAX,CMB} disagree with the 
predictions \cite{DurreretAl,LewandAlb} for  the cosmic microwave background
anisotropies based on topological defect models (see also ref.
\cite{DurreretAlRep}). This seems to point to the conclusion that if cosmic
strings were formed in the early universe they could not have been responsible
for structure formation. However, cosmic strings are still cosmologically
relevant and enjoy wide interest in cosmology.

   The most common field-theoretical model, which is used in order to
   describe the generation of cosmic strings during a phase transition, is the
   Abelian Higgs model. This model is defined by the action:

   \begin{equation}
   S = \int d^4 x \sqrt{\mid g\mid } \left({1\over 2}D_{\mu}\Phi ^
   {\ast}D^{\mu}\Phi
   - {{\lambda  }\over 4}(\Phi ^{\ast} \Phi - v^2)^2 -
   {1\over 4}{F}_{\mu \nu}{F}^{\mu \nu} + \frac{1}{16\pi G}
   {\cal R}\right)
   \label{higgsaction}
   \end{equation}
   where ${\cal R}$ is the Ricci scalar, ${F}_{\mu \nu}$ the Abelian field
   strength, $\Phi$ is a complex scalar field with vacuum expectation value
$v$
   and $D_\mu = \nabla _{\mu} - ieA_{\mu}$ is the usual gauge covariant
   derivative. We use units in which $\hbar=c=1$ and a "mostly minus" metric.

The simplest form of a cosmic string is just a direct generalization of the
 Nielsen-Olesen flux tube \cite{NO}, which takes into account its
gravitational
effect. Because of the cylindrical symmetry, we use a line element
   of the form:
   \begin{equation}
   ds^{2} = N^{2}(r)dt^{2} - d{r}^{2} - L^{2}(r)d{\varphi}^2 - K^{2}(r)dz^{2}
   \label{lineelement}
   \end{equation}
   and the usual Nielsen-Olesen ansatz for the +1 flux unit:
   \begin{eqnarray}
   \Phi=vf(r)e^{i\varphi} \hspace{10 mm} , \hspace{10 mm}
   A_\mu dx^\mu = {1\over e}(1-P(r))d\varphi
   \label{NOansatz}
   \end{eqnarray}

Since the flux tube is very concentrated along the symmetry axis, it is
quite natural
to expect that it will generate a spacetime geometry with the asymptotic
behavior
 of the general static and cylindrically-symmetric vacuum solution of
Einstein equations
   \cite{exact-sol}, the so-called Kasner solution:
   \begin{eqnarray}
   ds^{2} = (kr)^{2a}dt^{2} - (kr)^{2c}dz^{2} - dr^{2}
   - \beta^{2} (kr)^{2(b-1)}r^{2}d{\varphi}^2
   \label{Kasner1}
   \end{eqnarray}
   where $k$ sets the length scale while $\beta$ represents the asymptotic
   structure, as will be discussed below. The parameters $(a, b, c)$ must
satisfy the
Kasner conditions:
   \begin{equation}
   a + b + c=a^2 + b^2 + c^2=1
   \label{Kasner2}
   \end{equation}

More information about the solutions can be obtained by inspection of the
full system of
field equations with the appropriate energy-momentum tensor, which we will
not write here
(see however Section 4 or refs.\cite{vilsh,Garfinkle1,CLV}).
Here we will only
give a condensed summary. For
cylindrical  symmetry, the components of the energy-momentum tensor of the
 flux tube  ${\cal T}^\mu_\nu$  have the property of ${\cal T}^0_0={\cal
T}^{z}_{z}$ .
  This means that the solution will
   have a symmetry under boosts along the string  axis, i.e., $a=c$. The
Kasner
   conditions (\ref{Kasner2}) then  leave only two options, which are indeed
realized as
solutions of the full Einstein-Higgs system \cite{CLV}.

 The standard conic cosmic string solution \cite{Vil1,Garfinkle1} is
characterized  by
an asymptotic behavior given by (\ref{Kasner1}) with:
 \begin{equation}
   a=c=0 \;\;\; , \;\;\; b=1
   \label{SL}
   \end{equation}
 which is  evidently locally flat. In this case, the parameter $\beta$
represents a
   conic angular  deficit \cite{Marder2,Bonnor}, which is also related to
the mass
distribution  of the source.

In addition to the cosmic string solutions, there exists a second possibility:
  \begin{equation}
   a=c=2/3\;\;\; ,\;\;\;b=-1/3
   \label{MW}
   \end{equation}
   which is the same behavior as that of the Melvin solution \cite{Melvin}.
    Eq.(\ref{MW}) is therefore
referred to as the Melvin branch.
Note however that the magnetic field in this solution inherits the
exponential decrease (with $r$) of the original Nielsen-Olesen flux tube.
Therefore, it
is much more concentrated than
in the original Melvin solution where it decreases only according to a
power law.  
 
The main difference between the two branches lies in the Tolman mass,  
which is zero for the
conic cosmic string solutions, but non-zero for the Melvin solutions
\cite{CLV}. Moreover, the
central magnetic field is generally larger for the Melvin solutions.

The string-like solutions of the Abelian Higgs model are the simplest and
most studied
ones (see e.g. \cite{Garfinkle1}-\cite{CLV},
\cite{Linet2}-\cite{LagGarfinkle}).
 It is, however,  probable that
when high energy corrections are taken into account,
 gravity is not purely tensorial. A minimal modification suggested by
string theory
 \cite{GSW} is the introduction of a scalar degree of freedom, the dilaton
$\phi$,
 turning gravity into a scalar-tensor theory in the spirit of
Jordan-Brans-Dicke (JBD) theory. Studies of cosmic string solutions in the
framework
of JBD  theory
and its
extensions already exist in the literature \cite{GO, PM, Guimaraes}, and the
typical
characteristics of the string-like solutions are found to be quite
different from those
of the gauge strings of pure tensorial gravity. The main difference is the
absence
of asymptotically conic solutions due to the long range effect of the massless
JBD field. Unlike the Brans-Dicke
field, which is
 postulated not to couple to matter, the dilaton may do so and may further
change
the situation. This is represented by the following action \cite{GS,Santos},
which
 is written in the so called  "JBD/string frame \cite{FGN}":
 \begin{equation}
   S = \int d^4 x \sqrt{\mid g\mid } \left(e^{2a\phi}({ 1\over 2}D_{\mu}\Phi ^
   {\ast}D^{\mu}\Phi
   - {{\lambda  }\over 4}(\Phi ^{\ast} \Phi - v^2)^2  -
   {1\over 4}{F}_{\mu \nu}{F}^{\mu \nu})+
  \frac{1}{16\pi G} e^{-2\phi}(
   {\cal R}-4\nabla_{\mu}\phi \nabla^{\mu}\phi)\right)
   \label{higgsdilaction1}
   \end{equation}
The parameter $a$ serves as a general coupling constant between the dilaton
and
matter fields. Newton's constant is added explicitly in order to keep track
of the
dimensionality of the various fields. This makes the action in the
"Einstein/Pauli
frame \cite{FGN}" contain the standard Einstein-Hilbert Lagrangian:
\begin{eqnarray}
 S = \int d^4 x \sqrt{\mid g\mid } \left({1\over 2}e^{2(a+1)\phi}D_{\mu}\Phi ^
   {\ast}D^{\mu}\Phi
- {{\lambda  }\over 4} e^{2(a+2)\phi} (\Phi ^{\ast} \Phi- v^2)^2
-{1\over 4}e^{2a\phi}{F}_{\mu \nu}{F}^{\mu \nu}\right) \nonumber\\+
 \frac{1}{16\pi G}\int d^4 x \sqrt{\mid g\mid }
({\cal R}+2\nabla_{\mu}\phi \nabla^{\mu}\phi )
   \label{higgsdilaction2}
   \end{eqnarray}
It is related to the former by the metric redefinition:
 \begin{equation}
   {g}^{(s)}_{\mu \nu} =   e^{2\phi} {g}^{(E)}_{\mu \nu}
   \label{EJBD}
   \end{equation}
where $ {g}^{(s)}_{\mu \nu}$ is the metric tensor used in
(\ref{higgsdilaction1}) and
$ {g}^{(E)}_{\mu \nu}$ is used in (\ref{higgsdilaction2}). From now on we
will use the
Einstein frame for a clearer physical picture.

This system was studied by Gregory and Santos \cite{GS}. Their
 analysis was done to first order in the parameter $\gamma=8\pi Gv^2$ and
is therefore
valid only for weak gravitational fields. They concentrated in the one
branch which contains for $a=-1$ asymptotically conic solutions. However,
since for
$a\neq -1$ no asymptotically conic solutions exist, there is no natural
criterium
to prefer one branch over the other, even if one is "more flat" than the
other. Moreover,
the weak-field approximation breaks down both near the core of the string
and far away from the string thus calling for a further investigation of
the system.

In this paper, we expand the study of cosmic strings in dilaton gravity in two
directions.\\
1)We analyze both branches and do not limit ourselves to weak gravitational
fields.\\
2)We modify the dilaton coupling to matter fields in such a way that it
will be uniform
and the original Einstein-Higgs system  (\ref{higgsaction}) is recovered in
a certain limit.

\section{Field Equations and Asymptotic Solutions}
\setcounter{equation}{0}
   \label{secAsympt}

The field equations for the Einstein-Higgs-dilaton action
(\ref{higgsdilaction2})
are straightforward to obtain and we give them already for flux tubes with
cylindrical
symmetry. Their structure becomes more transparent if we use dimensionless
quantities.  As a length scale we use $1/\sqrt{\lambda v^2}$ (the
``correlation
   length" in the superconductivity terminology). We therefore change to the
   dimensionless length
   coordinate $x=\sqrt{\lambda v^2}r$ and we use the metric component
   $\sqrt{\lambda
   v^2}L$ which we still denote by $L$. We also introduce the "Bogomolnyi
parameter"
   $\alpha=e^2/\lambda$ in addition to $\gamma=8\pi Gv^2$, which has been
already
defined above. In terms of these new quantities we
   get, for fixed $a$, a two parameter system of five coupled non-linear
ordinary
differential equations (the prime denotes $d/dx$):
\begin{eqnarray}
   \frac{(e^{2(a+1)\phi}\;N^2 Lf')'}{N^2 L} +
   \left( e^{2(a+2)\phi}\;(1-f^2) -
   e^{2(a+1)\phi}\;\frac{P^2}{L^2}\right)f = 0
  \label{eqf}
   \end{eqnarray}
   \begin{eqnarray}
   \frac{L}{N^2}\left( e^{2a\phi}\;\frac{N^2 P'}{L}\right)' -
   \alpha e^{2(a+1)\phi}\;f^2 P = 0
  \label{eqP}
   \end{eqnarray}
   \begin{eqnarray}
   \frac{(LNN')'}{N^2 L} &=&\gamma\left(
   e^{2a\phi}\;\frac{P'^2}{2\alpha L^2} - \frac{1}{4}
   e^{2(a+2)\phi}\;(1-f^2)^2\right)
  \label{eqN}
   \end{eqnarray}
   \begin{eqnarray}
   \frac{(N^2 L')'}{N^2 L}
   &=& - \gamma\left( e^{2a\phi}\;\frac{P'^2}{2\alpha
   L^2} + e^{2(a+1)\phi}\;\frac{P^2 f^2}{L^2} + \frac{1}{4}
   e^{2(a+2)\phi}\;(1-f^2)^2\right)
  \label{eqL}
   \end{eqnarray}
   \begin{eqnarray}
   \frac{(N^2 L\phi')'}{N^2 L}
  = \gamma\left(
  \frac{a+1}{2}e^{2(a+1)\phi}\;\left( \frac{f^2 P^2}{ L^2}
  + f'^2 \right)+ae^{2a\phi}\frac{P'^2}{2\alpha L^2}
   + \frac{a+2}{4} e^{2(a+2)\phi}\;(1-f^2)^2
   \right)
  \label{eqphi}
   \end{eqnarray}
   The sixth function, the metric component $K(x)$, turns out to be equal to
$N(x)$ due to the high symmetry. We have also to keep in mind the existence
of the constraint
   which comes from the
   ($rr$) Einstein equation, and gets the  following form:
   \begin{eqnarray}
   \frac{N'}{N} \left(2\frac{L'}{L}+ \frac{N'}{N}\right)
  &=& \nonumber\\ \phi'^2 &+&
   \gamma\left( e^{2(a+1)\phi}\;\frac{f'^2}{2} +
   e^{2a\phi}\;\frac{P'^2}{2\alpha L^2} -
   e^{2(a+1)\phi}\;\frac{P^2 f^2}{2L^2} -
   \frac{1}{4}e^{2(a+2)\phi}\;(1-f^2)^2 \right)
  \label{constraint}
   \end{eqnarray}

The field equations are supplemented by the following boundary conditions,
that
should be satisfied by the scalar and gauge fields:

   \begin{eqnarray}
   f(0)=0 &, & \lim_{x\rightarrow \infty} f(x) = 1 \nonumber \\
   P(0)=1 &, & \lim_{x\rightarrow \infty} P(x) = 0
   \label{boundarycond}
   \end{eqnarray}
   These are just the usual flux tube boundary conditions borrowed from the
flat
space version. Moreover, regularity of the geometry on the symmetry axis $x=0$
will be guaranteed by the "initial conditions":
   \begin{eqnarray}
   L(0)=0 &, & L'(0) = 1 \nonumber \\
   N(0) = 1 &, & N'(0) = 0
   \label{initcond}
   \end{eqnarray}
The presence of the dilaton requires additional conditions, which we impose on
 the dilaton field on the axis:
  \begin{eqnarray}
  \phi(0)=0 &, & \phi'(0) = 0
   \label{dilatoninitcond}
   \end{eqnarray}
The condition $\phi'(0) = 0$ follows directly from the equations of motion
and the constraint (\ref{constraint}), using also
eqs.(\ref{boundarycond})-(\ref{initcond}). As for the condition
$\phi(0)=0$, notice
that the equations of motion (2.1)-(2.6) are invariant under the following
transformation:
\begin{eqnarray}
\phi&\rightarrow&\phi +\phi_0 \nonumber\\
x&\rightarrow&xe^{-\phi_0} \nonumber\\
L&\rightarrow&Le^{-\phi_0} \nonumber\\
\gamma &\rightarrow& \gamma e^{-2(a+1)\phi_0}
\end{eqnarray}
where $\phi_0$ is a constant. Thus, we can trivially get the solution for
$\phi(0)=\phi_0$ from the solution for $\phi(0)=0$ by simple rescaling.

\begin{figure}[!t]
   \begin{center}
   \includegraphics[width=10cm]{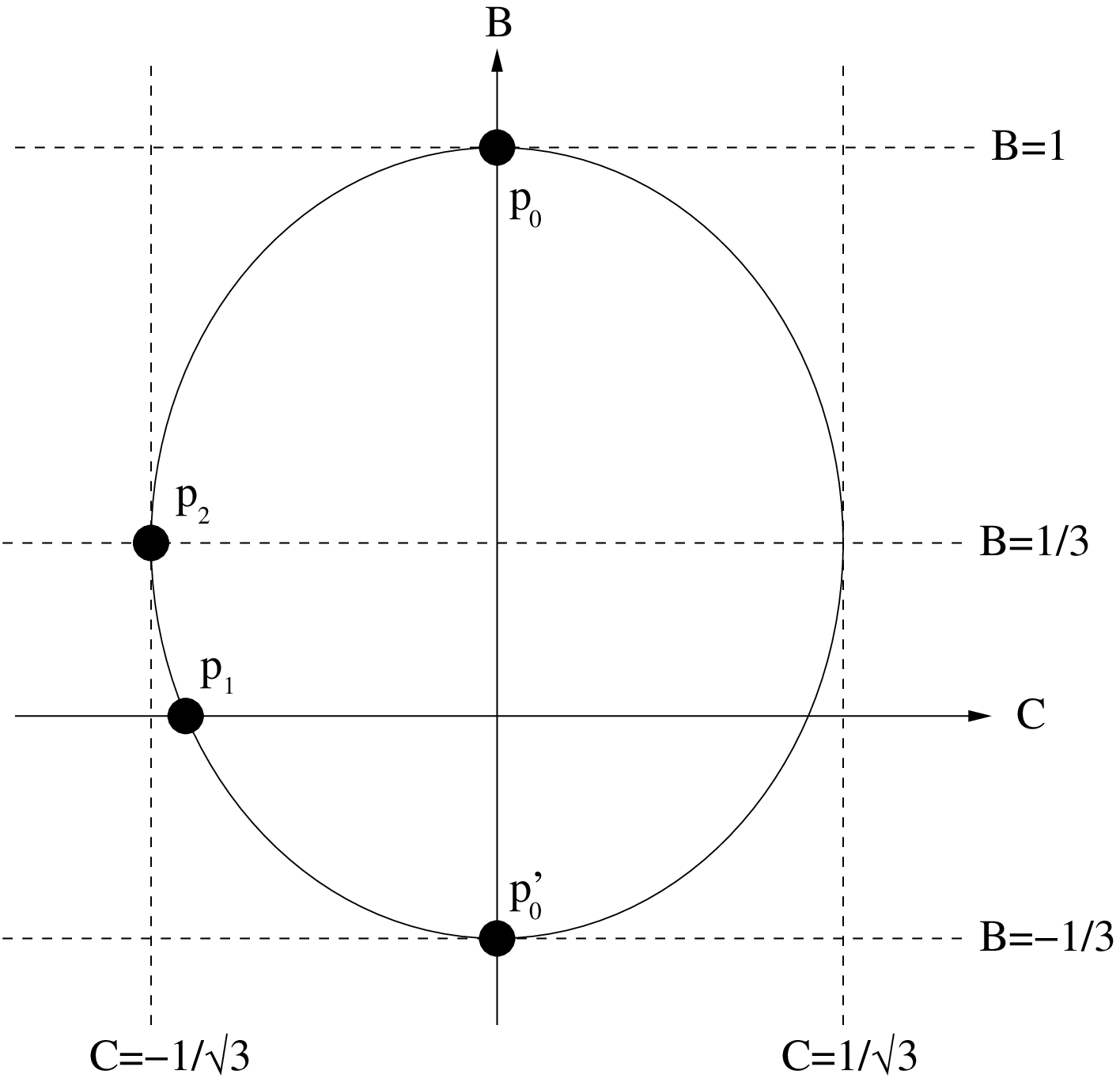} \\
   \caption{The ellipse - eq.(\ref{elipse}). The origin is a focal point.
The special points marked in the figure correspond to solutions discussed
in the paper. }
 \label{ellipsea0}
   \end{center}
   \end{figure}

Since we are looking for string-like solutions to the system, we may easily
get
the  asymptotic behavior of the metric components from the assumption that the
 right-hand sides of the Einstein equations vanish exponentially fast, as
in the
 case of pure tensorial gravity. Consequently, the
system reduces to a simple set of equations, which are easily integrated to
power
law solutions:
 \begin{eqnarray}
   N(x)\sim \kappa x^A \hspace{6 mm},\hspace{6 mm} L(x)\sim \beta x^B
\hspace{6 mm},\hspace{6 mm} e^{\phi(x)}\sim \delta x^C
   \label{asymptvac}
   \end{eqnarray}
The 3 parameters $A$, $B$ and $C$ are subjected to two conditions, which
may be
viewed as a
generalization of the Kasner conditions, eq.(\ref{Kasner2}):
\begin{equation}
  2 A + B=1 \hspace{6 mm},\hspace{6 mm} A^2  + 2AB =C^2
   \label{dilatonKasner}
   \end{equation}
These two conditions leave a one-parameter family of solutions, which may be
visualized by the ellipse in the $B-C$ (or equivalently $A-C$) plane
(fig.\ref{ellipsea0}):
\begin{eqnarray}
  \frac{9}{4}(B-\frac{1}{3})^2+3C^2 = 1
\label{elipse}
   \end{eqnarray}
The 2 branches of the Einstein-Higgs system (the cosmic string branch and the
Melvin branch) are represented in this picture by the top ($p_0$) and
bottom ($p'_0$)
points of the ellipse, respectively. These points correspond to an
asymptotically
constant dilaton ($C=0$), and as we will show later, only the first is
realized as
an actual solution for the special value of the coupling constant $a=-1$.

 The parameters $A$, $B$ and $C$ are related to the matter distribution
 through the following
 three quantities: The Tolman mass (per unit length) $M$, or rather its
 dimensionless representative $m=GM$:
 \begin{equation}
m=\frac{\gamma}{2} \int _{0}^\infty
dx\;  N^2 L \left(e^{2a\phi}\;\frac{P'^2}{2\alpha L^2} - \frac{1}{4}
  e^{2(a+2)\phi}\;(1-f^2)^2\right)=\frac{1}{2} \lim_{x\rightarrow\infty} (LNN')
\label{mass}
\end{equation}
and two others:
\begin{eqnarray}
w=\frac{\gamma}{2} \int _{0}^\infty
dx\;  N^2 L \left( e^{2a\phi}\;\frac{P'^2}{2\alpha
   L^2} + e^{2(a+1)\phi}\;\frac{P^2 f^2}{L^2} + \frac{1}{4}
   e^{2(a+2)\phi}\;(1-f^2)^2\right)=\nonumber\\
   -\frac{1}{2} (\lim_{x\rightarrow\infty} (N^2 L')-1)
\label{ANGULARW}
\end{eqnarray}
which is related to the string angular deficit, and:
\begin{eqnarray}
D=\gamma \int _{0}^\infty
dx\;  N^2 L \left( \frac{a+1}{2}e^{2(a+1)\phi}\;\left( \frac{f^2 P^2}{ L^2}
  + f'^2 \right)+ae^{2a\phi}\frac{P'^2}{2\alpha L^2}
   + \frac{a+2}{4} e^{2(a+2)\phi}\;(1-f^2)^2 \right) \nonumber\\
   = \lim_{x\rightarrow\infty} (N^2 L\phi ')
   \label{DilatonCh}
\end{eqnarray}
which may be interpreted as a dilaton charge. Note that $w$ is 
manifestly positive definite and $D$ is positive definite for $a\geq 0$.  
The Tolman mass is also non-negative since it is proportional to the power  
$A$ which is non-negative due to (\ref{dilatonKasner}). 
We further notice that, as in the non-dilatonic case \cite{CLV},  
$w$ may be expressed in terms of the mass parameter $m$ and the  
magnetic field on the axis, which we represent by the dimensionless  
parameter ${\cal B}$: 
 
\begin{equation}
w=\frac{1}{2} {\cal B} -m\;,\;\;
 \mbox{where} \;\;\; {\cal B}= -\frac{\gamma}{\alpha}\lim_{x\rightarrow0}\left(
\frac{P'(x)}{L(x)} \right)
\label{centralmagn}
\end{equation}
The parameters $A$, $B$ and $C$ are easily found to be
expressible in terms of $m$, $D$ and $w$ or as:
\begin{eqnarray}
A=\frac{2m}{6m+1-{\cal B}} ,\hspace{6 mm}
B=\frac{2m+1-{\cal B}}{6m+1-{\cal B}} ,\hspace{6 mm}
C=\frac{D}{6m+1-{\cal B}}.
\label{sourceAndpowers}
\end{eqnarray}
The first of the two relations (\ref{dilatonKasner}) is satisfied identically;
the second translates in terms of $m$, $D$ and $\cal B$ to:
\begin{equation}
D^2=4m(3m+1-{\cal B}).
\label{dilatonKasnerSource}
\end{equation}

\section{String-like Solutions}
\begin{figure}[!t]
   \begin{center}
   \includegraphics[width=7.8cm,angle=270]{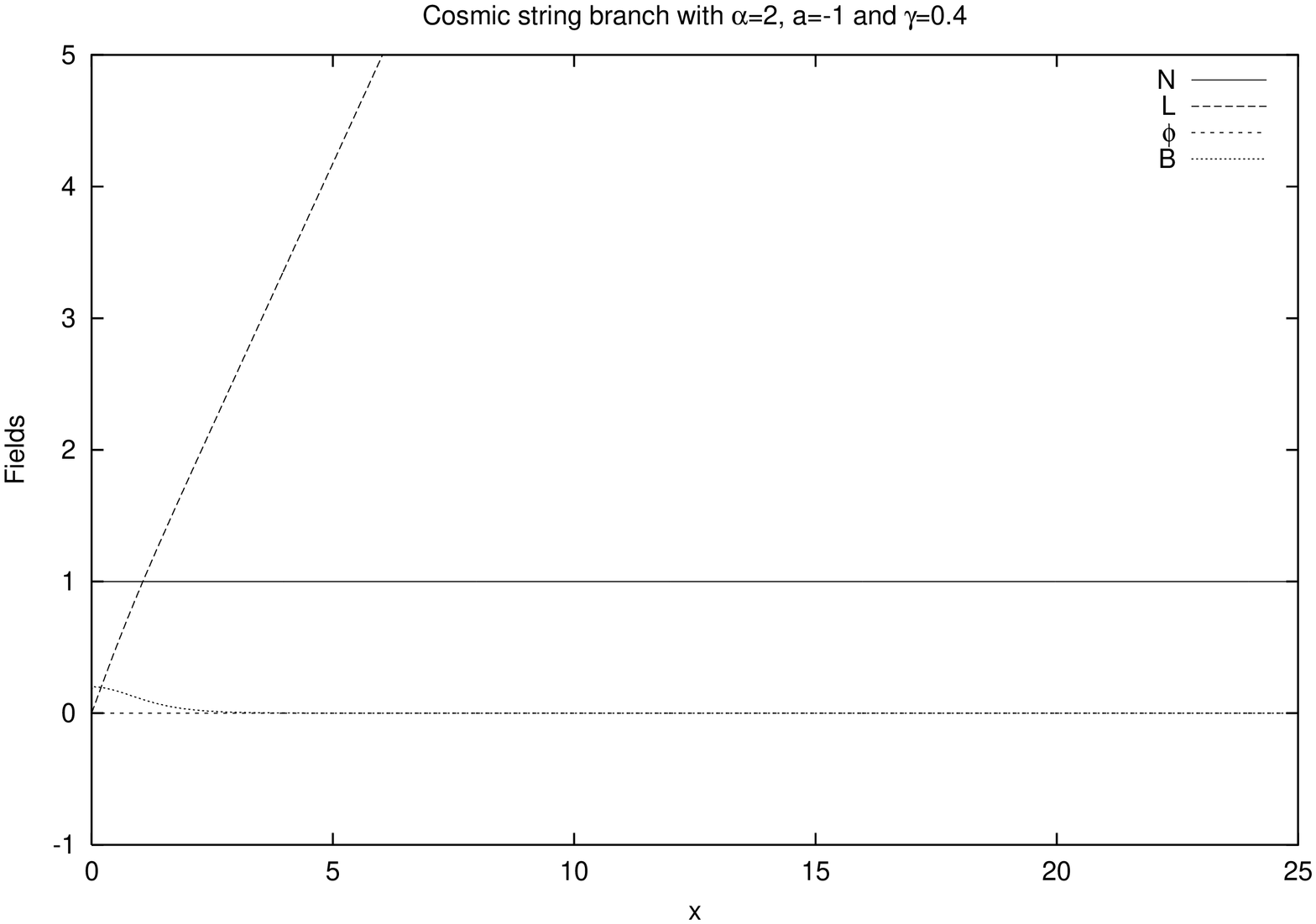}\\(a)\\
    \includegraphics[width=7.8cm,angle=270]{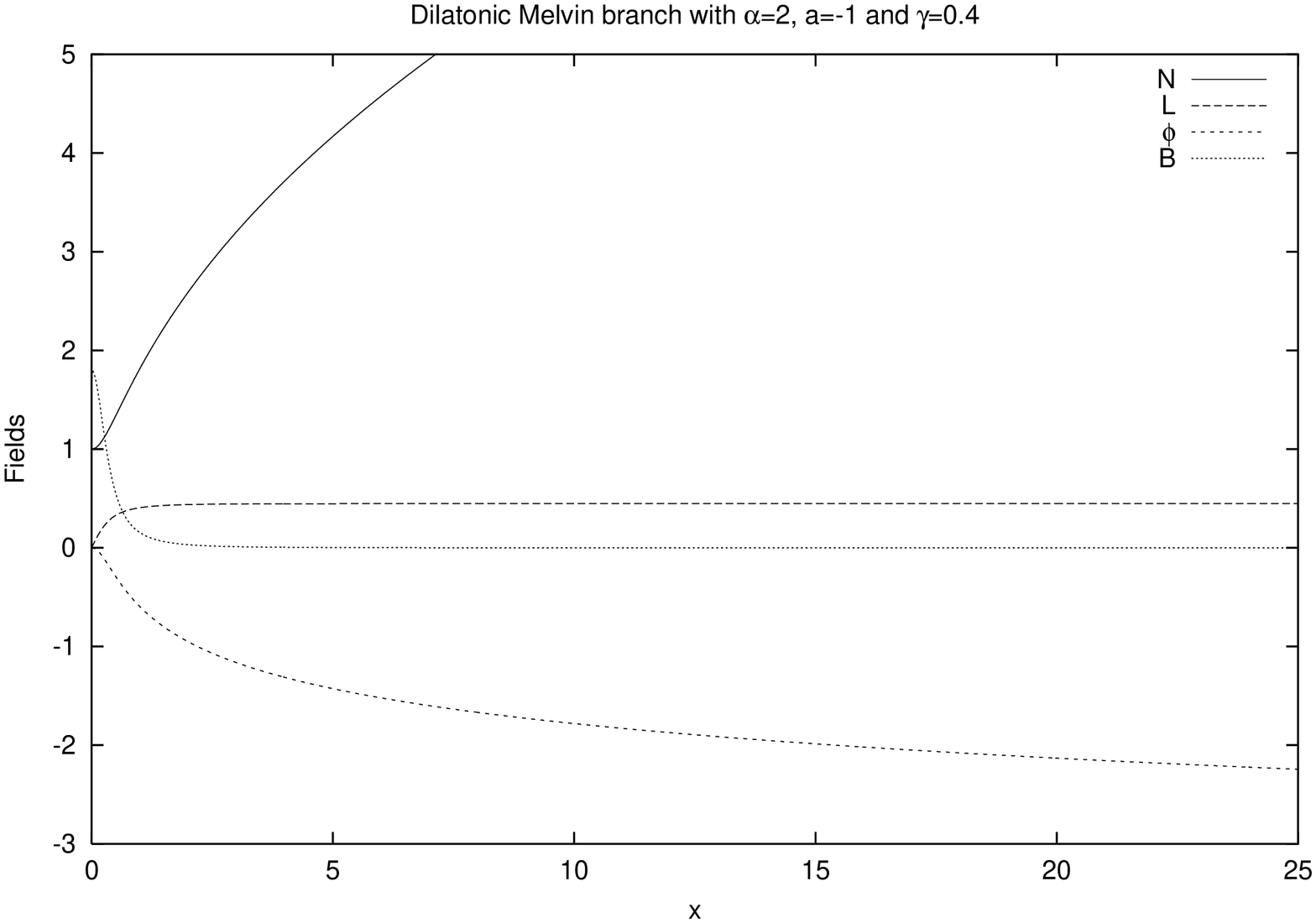}\\(b)
\caption{The solutions for $a=-1$, $\alpha = 2$, $\gamma = 0.4$.
(a) Cosmic string branch. (b) Dilatonic Melvin branch. }
\label{Soltna2a-1g0.4}
   \end{center}
   \end{figure}

\begin{figure}[!t]
   \begin{center}
   \includegraphics[width=7.8cm,angle=270]{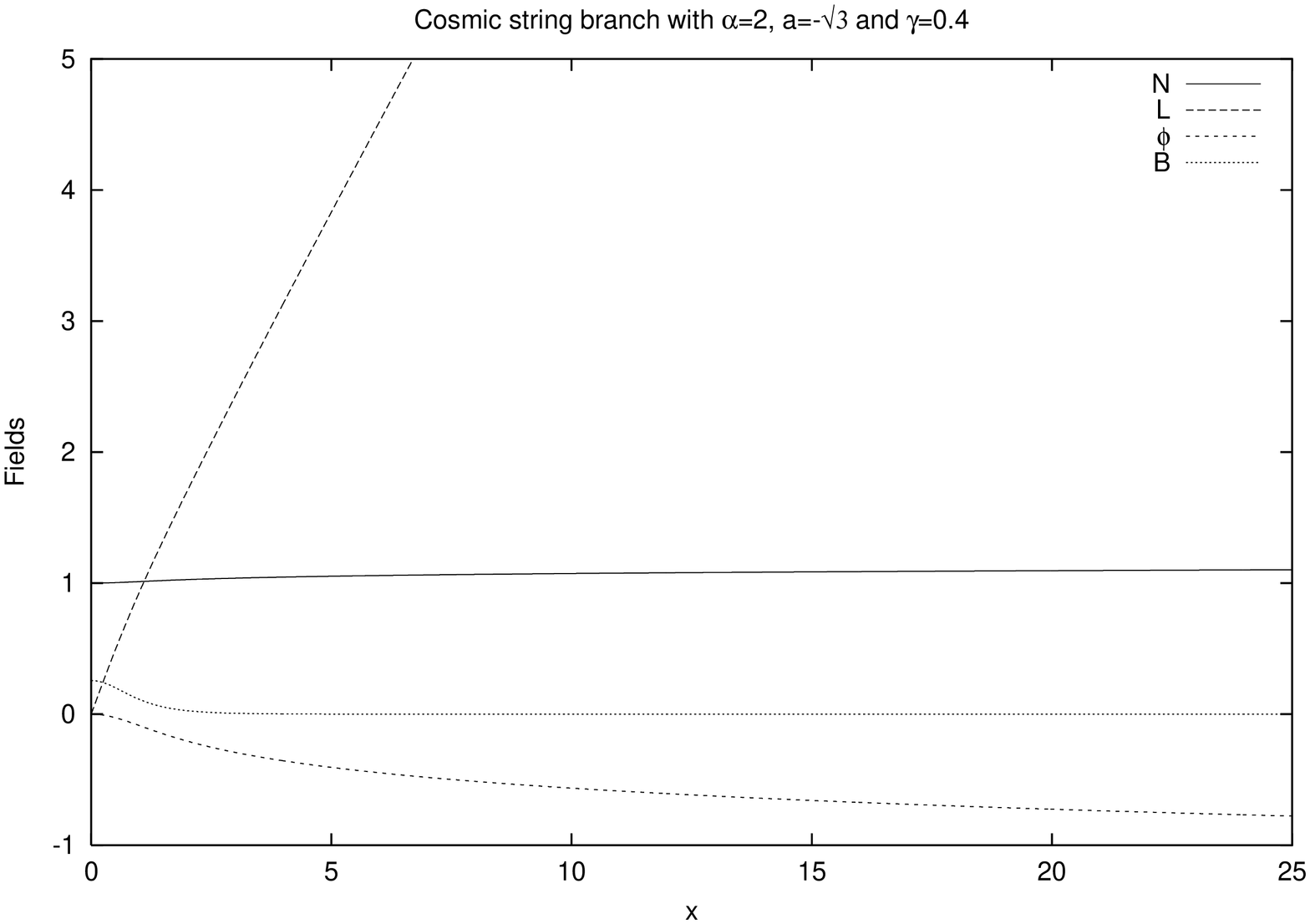}\\(a)\\
    \includegraphics[width=7.8cm,angle=270]{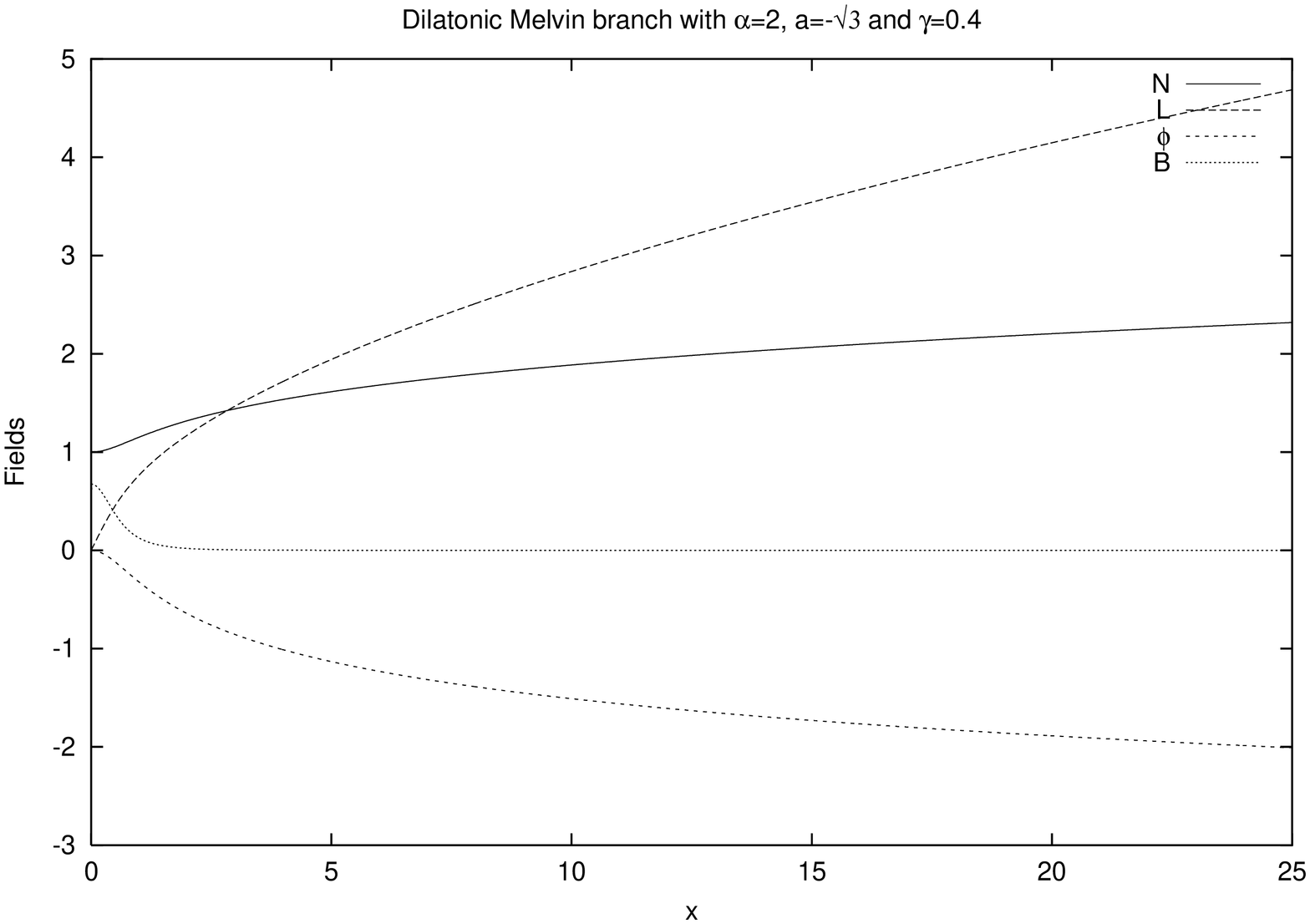}\\(b)
\caption{The solutions for $a=-\sqrt{3}$, $\alpha = 2$, $\gamma = 0.4$.
(a) Cosmic string branch. (b) Dilatonic Melvin branch. }
\label{Soltna2a-s3g0.4}
   \end{center}
   \end{figure}

\begin{figure}[!t]
   \begin{center}
   \includegraphics[width=7.8cm,angle=270]{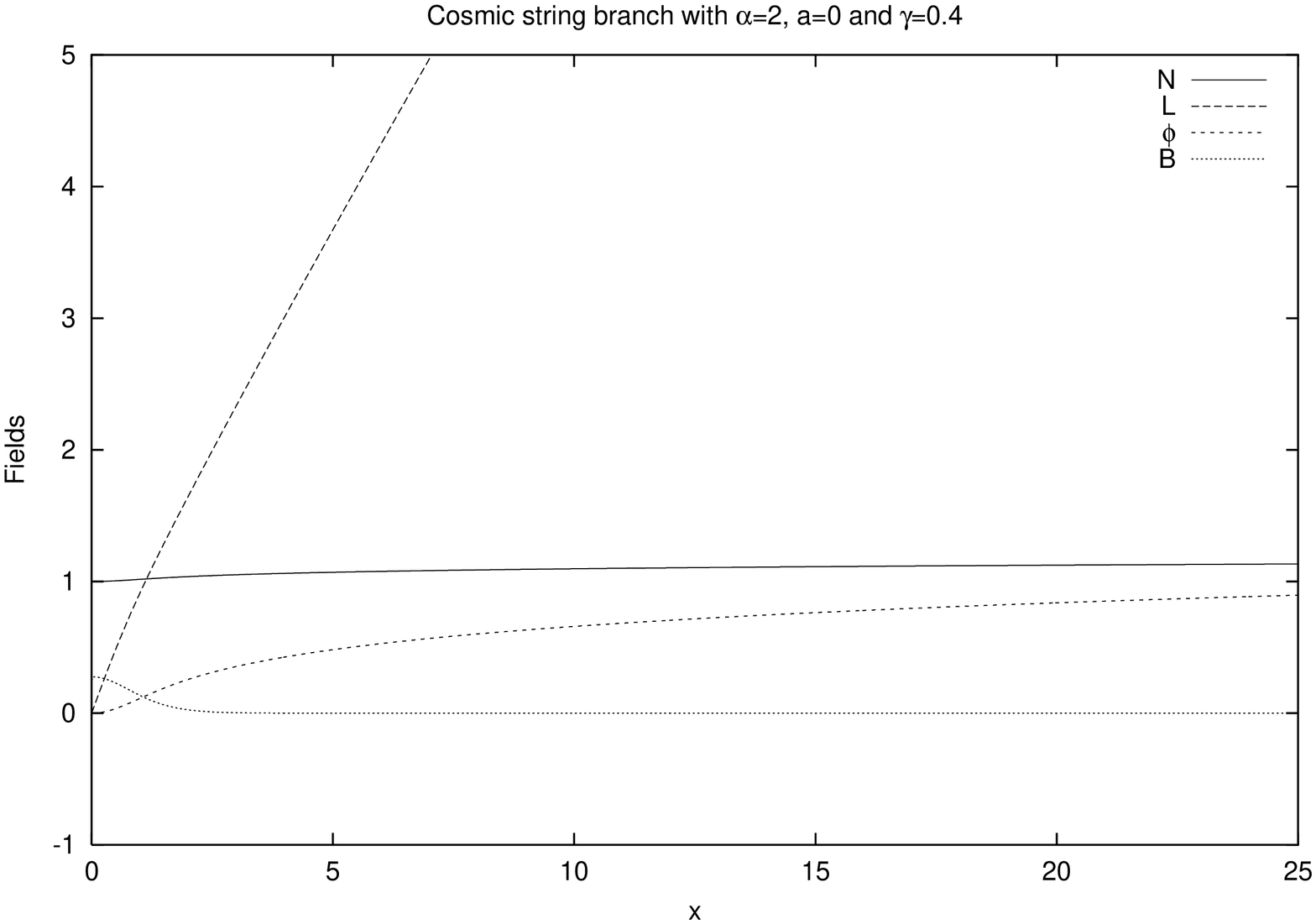}\\(a)\\
    \includegraphics[width=7.8cm,angle=270]{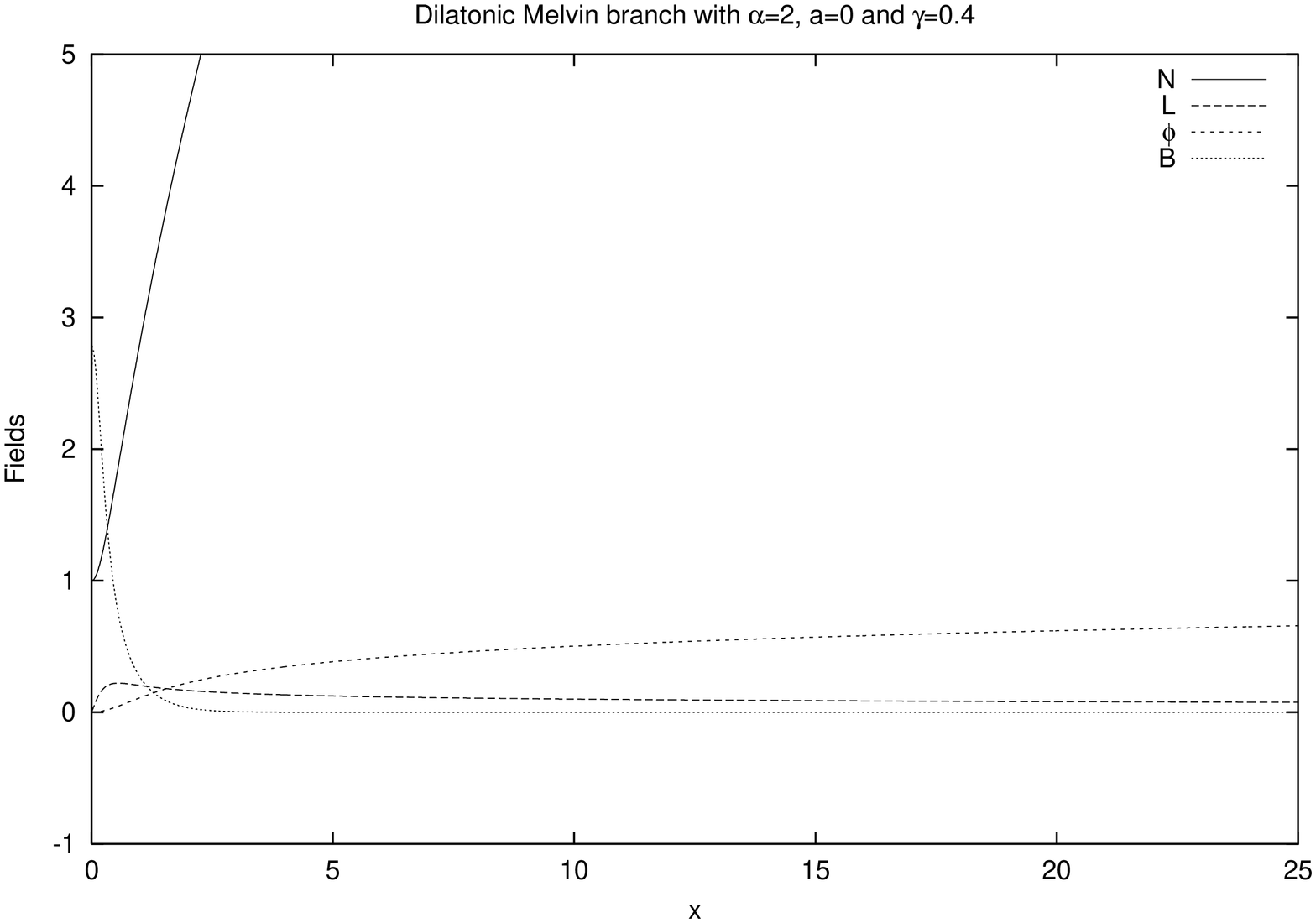}\\(b)
\caption{The solutions for $a=0$, $\alpha = 2$, $\gamma = 0.4$.
(a) Cosmic string branch. (b) Dilatonic Melvin branch. }
\label{Soltna2a0g0.4}
   \end{center}
   \end{figure}

\setcounter{equation}{0}
\label{secSols}
Regarding the complexity and non-linearity of the system, it is quite
evident that
a detailed structure of the solutions can be obtained only numerically.
We have performed a numerical analysis of the system and found that the
solutions
are paired, as in the non-dilatonic case, and have the characteristics that
will be
described below.
The solutions have been found numerically by first discretizing the radial
coordinate and then applying a relaxation procedure to the set of
non-linear and
coupled algebraic equations, ${\cal F} (\bar f)= 0$, in the set of scalar
functions evaluated at the grid points, $\bar f$. If the initial guess
$\bar f_i$
is good enough then the iteration obtained by
$\bar f_{i+1} = \bar f_i + \Delta \bar f_i$,  where $\Delta f_i$
is found by solving the linear equation:
\begin{equation}
\nabla {\cal F} (\bar f_i) \, \Delta \bar f_i \: = \: - {\cal F} (\bar f_i)
\end{equation}
(Gauss elimination with scaling and pivoting will do), will converge towards a
solution for the discretized variables. We constructed a starting guess
satisfying the boundary and initial conditions and still having several free
parameters. Changing gradually these parameters, we eventually stumbled into
the
region of convergence and  a  solution was found. Having obtained one solution
of this type, the rest could easily be generated by moving around in the
parameter
space in sufficiently small steps.
Representative plots of the solutions are given in figs.\ref{Soltna2a-1g0.4},
 \ref{Soltna2a-s3g0.4} and \ref{Soltna2a0g0.4}.

The simplest picture emerges in the case $a=-1$.
The two branches of solutions for this paticular value of $a$ are particularly
interesting. For fixed $\alpha$, both solutions exist up to some critical
$\gamma$;
see fig.5. One is an ordinary cosmic string, as in
Einstein gravity with the  additional feature of  a non-constant dilaton. This
solution has $A=0$, $B=1$, $C=0$ independent of  $\alpha$ and $\gamma$. The
metric
is very similar  to the usual gauge string,  but there are small  deviations,
due to the presence of the dilaton. It becomes identical to the metric of the
Einstein gravity gauge string only in the Bogomolnyi limit ($\alpha=2$),
where eqs.(\ref{eqf})-(\ref{constraint}) reduce to:
\begin{eqnarray}
P'=L(f^2-1) \\
f'=\frac{Pf}{L} \\
L'-\frac{\gamma}{2} P(1-f^2) = 1-\frac{\gamma}{2}
\label{Bog}
\end{eqnarray}
with $N=1$ and $\phi=0$, i.e. first order differential equations for $L$, $f$ 
 and $P$. A typical solution of this kind is depicted in  
fig.\ref{Soltna2a-1g0.4}(a).  
 
The second branch in the case $a=-1$ has for all values of  $\alpha$ and 
$\gamma$ below the critical curve (fig.5), $C=-\frac{1}{2}$ and  $B=0 $.
Accordingly,
$A=\frac{1}{2}$. This solution is quite different from its non-dilatonic
analog, the main difference being the asymptotic spatial geometry which
is now cylindrical.

At the critical curve (fig.5), the two types of solutions become identical.
Asymptotically the corresponding spatial geometry is conic with deficit angle
$\delta\varphi=2\pi$. The critical curve itself can be approximated by a
power-law:
\begin{equation}
\gamma \approx c_1 \alpha^{c_2}
\end{equation}
where $c_1\approx 1.66$ and $c_2\approx 0.275$. This is quite, although not
exactly,
similar to the result obtained in tensorial gravity \cite{CLV}.

 The
$\alpha$ and
$\gamma$ - independence of
$A$,
$B$ and
$C$ for
$a=-1$ is however the only case of such behavior. The generic one is such
that for
a given value of
$a$, the powers of the asymptotic
form of the metric components and the dilaton change with $\alpha$ and
$\gamma$.
Put differently, we have now a situation where the values of $A$, $B$ and
$C$ depend
upon the
three parameters $a$, $\alpha$ and $\gamma$. Therefore, the universal
asymptotic form of the metric tensor, which corresponds to the two extremal
points
on the ellipse, is now replaced by a motion along the ellipse whose
specific details
depend on the  value of  $a$.  Additional values of $a$ were studied in
detail and
here we give few further results for two more special representative points,
so in
all there are:
 \begin{equation}
 a= -\sqrt{3},\hspace{3 mm}-1,\hspace{3 mm} 0
   \label{avalues}
 \end{equation}
These values were selected for the following reasons: Concentrating on the
Einstein-dilaton-Maxwell system, $a=-1$ corresponds to certain compactified
superstring
theories \cite{sen}, $a=0$ corresponds to Jordan-Brans-Dicke theory and
$a=-\sqrt{3}$
corresponds to ordinary Kaluza-Klein theory.

In both cases $a= -\sqrt{3}$, $a=0$  the same two-branch picture emerges. 
One branch
always contains the upper point of the ellipse ($p_0$) at the limit
$\gamma \rightarrow  0$, such that for
small $\gamma$ the "almost asymptotically conic" solutions of Gregory and
Santos
 \cite{GS} are obtained. For large $\gamma$ the deviations from
asymptotically conic
geometry are more pronounced. At the same time a gravitational Newtonian
potential
appears, so this kind of string exerts force on non-relativistic test
particles far
outside its core.

The other branch exhibits also an $\alpha$ and $\gamma$ dependence of the
powers
$A$, $B$ and $C$, but has its $\gamma \rightarrow  0$ limit at
$a$-dependent points. This $a$-dependence can be understood if we note that
the  $\gamma \rightarrow  0$
limit corresponds to two types of asymptotic behavior. One is the
asymptotically conic
cosmic string behavior and the other is that of the dilatonic Melvin
universe
\cite{Santos,DGKT}, which in our coordinate system corresponds to an
asymptotic behavior with the following powers:
\begin{equation}
   A =\frac{2}{a^2+3}\; , \hspace{6 mm}  B =\frac{a^2-1}{a^2+3} \; ,
\hspace{6 mm} C =\frac{2a}{a^2+3}
   \label{dilatonMelvin}
   \end{equation}

The situation therefore is the following. First we fix $a$. One branch (the
"upper"
or  "cosmic string" branch) has the point $p_0$, i.e., $A=C=0$, $B=1$ as a
starting
point for $\gamma \rightarrow  0$, and as $\alpha$ and $\gamma$ change, the
powers
$B$  and $C$ move  along the ellipse  while $A$ changes accordingly (see
\ref{dilatonKasner}).  The other  branch  (the "lower" or "dilatonic
Melvin" branch)
starts, for   $\gamma \rightarrow  0$, with  the above values
eq.(\ref{dilatonMelvin}) for the powers $A$, $B$ and  $C$ and the  motion along
the  ellipse is done in the opposite direction so the two  branches
approach each
other and "meet" at a point in between. In the  meeting point, the
solutions not
only have the same asymptotic behavior, but they become  actually identical.

There are no asymptotically conic solutions except for $a=-1$. Actually,
this property
can be understood analytically. For this, we note that a useful relation is
obtained by
adding  the equation for the dilaton, (\ref{eqphi}) to the equation for the
metric
component $N$, (\ref{eqN}) and integrating from $0$ to $x$, using
the boundary conditions (on the axis):
\begin{eqnarray}
  & & \mbox{}  N^2 L(\phi' + N'/N) = \nonumber\\
& & \mbox{} \gamma (a+1)\int_{0}^{x} d\bar{x}  N^2 L \left(
  \frac{1}{2}e^{2(a+1)\phi}\;\left( \frac{f^2 P^2}{ L^2}
  + f'^2 \right) + e^{2a\phi}\frac{P'^2}{2\alpha L^2}
   + \frac{1}{4} e^{2(a+2)\phi}\;(1-f^2)^2
   \right)
  \label{InteqNphi}
 \end{eqnarray}
where we denote by $\bar{x}$ the integration variable. In the case $a=-1$,
this equation
 is easily integrated again and gives in this case:
\begin{equation}
   N = e^{-\phi}
   \label{Nandphi}
   \end{equation}
It follows that the dilaton charge is minus twice the Tolman mass, $D=-2m$.  
This imposes an additional condition on the powers of the asymptotic metric  
components for any $\alpha$, $\gamma$ which selects two points on the  
ellipse. The first is: 
 \begin{equation} 
   A=C=0 \;\;\; , \;\;\; B=1 
   \label{DilConic} 
   \end{equation} 
which is  evidently locally flat.  
 
It is also easy to prove the converse, i.e., that there are no asymptotically  
conic solutions for $a\neq -1$. In order to do it, we start again with  
eq.(\ref{InteqNphi}) and notice also that for asymptotically conic solutions  
we need $A=0$ and $B=1$, which yield $C=0$. Thus, $N'/N$ and $\phi'$ vanish  
asymptotically faster than $1/x$ and 
the left-hand side of eq.(\ref{InteqNphi}) vanishes asymptotically. But the  
right hand side of this equation is an integral of a sum of 4  
positive-definite terms and it can vanish only if $a=-1$. 
 
It also follows that in this asymptotically conic solutions $D=m=0$, while  
the central magnetic field is given by: 
\begin{equation} 
{\cal B}=1-\kappa ^2\beta=1-\kappa ^2 \left( 1 - \frac{\delta 
\varphi}{2\pi}\right) 
\end{equation} 
where $\delta \varphi$ is the deficit angle. 
 
There is as usual a second point on the ellipse namely: 
   \begin{equation} 
   A=-C=1/2 \;\;\; ,\;\;\;B=0 
   \label{DilMelvin} 
   \end{equation} 
As indicated above, this represents the same asymptotic behavior as the  
dilatonic Melvin solution (with $a=-1$). This solution has non-vanishing  
dilaton charge and Tolman mass. The central magnetic field fulfills the  
equation: 
\begin{equation} 
{\cal B}=1+2m 
\end{equation} 

 \section{Uniform Dilaton Coupling}
\setcounter{equation}{0}
   \label{secUnif}

The model analyzed so far has a dilaton which couples (in the
Einstein/Pauli frame)
with different strengths to various matter fields. As a consequence, we do
not return
to the pure tensorial (Einstein) gravity in the limit of vanishing dilaton
coupling.
The only case of identical solutions of the two systems is the cosmic
string solutions
 with $a=-1$ and $\alpha  =2$. It seems
therefore natural to consider a uniform coupling of the matter fields to  the
dilaton.  This uniform coupling will be described by the following action:
\begin{equation}
   S = \int d^4 x \sqrt{\mid g\mid } \left(e^{2a\phi} \left({ 1\over
2}D_{\mu}\Phi ^
   {\ast}D^{\mu}\Phi
   - {{\lambda  }\over 4}(\Phi ^{\ast} \Phi - v^2)^2  -
   {1\over 4}{F}_{\mu \nu}{F}^{\mu \nu}\right)+
  \frac{1}{16\pi G}( {\cal R}+2\nabla_{\mu}\phi \nabla^{\mu}\phi)\right)
   \label{uniformaction}
   \end{equation}

The corresponding field equations for the cylindrically symmetric case will
be a simplified version of (\ref{eqf})-(\ref{constraint}):
\begin{eqnarray}
   \frac{(e^{2a\phi}\;N^2 Lf')'}{N^2 L} +
   e^{2a\phi}\; \left((1-f^2) - \frac{P^2}{L^2}\right)f = 0
  \label{eqfu}
   \end{eqnarray}
   \begin{eqnarray}
   \frac{L}{N^2}\left( e^{2a\phi}\;\frac{N^2 P'}{L}\right)' -
   \alpha e^{2a\phi}\;f^2 P = 0
  \label{eqPu}
   \end{eqnarray}
   \begin{eqnarray}
   \frac{(LNN')'}{N^2 L} &=&\gamma e^{2a\phi}\;\left(
   \frac{P'^2}{2\alpha L^2} - \frac{1}{4}(1-f^2)^2\right)
  \label{eqNu}
   \end{eqnarray}
   \begin{eqnarray}
   \frac{(N^2 L')'}{N^2 L}
   &=& - \gamma\ e^{2a\phi}\;\left(\frac{P'^2}{2\alpha
   L^2} + \frac{P^2 f^2}{L^2} + \frac{1}{4} (1-f^2)^2\right)
  \label{eqLu}
   \end{eqnarray}
   \begin{eqnarray}
   \frac{(N^2 L\phi')'}{N^2 L}
  = \gamma a e^{2a\phi}\;\left(
  \frac{1}{2}\left( \frac{f^2 P^2}{ L^2}
  + f'^2 \right)+\frac{P'^2}{2\alpha L^2}
   + \frac{1}{4} \;(1-f^2)^2
   \right)
  \label{eqphiu}
   \end{eqnarray}
  with the constraint:
   \begin{eqnarray}
   \frac{N'}{N} \left(2\frac{L'}{L}+ \frac{N'}{N}\right)
  &=& \nonumber\\\phi'^2 &+&
   \gamma\ e^{2a\phi}\;\left( \frac{f'^2}{2} +
   \frac{P'^2}{2\alpha L^2} -
  \frac{P^2 f^2}{2L^2} -
   \frac{1}{4}(1-f^2)^2 \right)
  \label{constraintu}
   \end{eqnarray}

The asymptotic behavior of the metric components and the dilaton field are
obtained as before
and the results are identical to (\ref{asymptvac})-(\ref{elipse}).

Analysis of the new system shows that the solutions do not change much except
in  the case $a=-1$, where an asymptotically conic solution does not exist any
more, see fig.6.  Actually, there are no asymptotically conic solutions in this
system for any combination of  parameters and in this respect the solutions
are more
similar to the string-like solutions in  JBD theory. It is quite simple to show
that the only case where asymptotically conic solutions  exist, in the case of
this uniform coupling, is $a=0$. The way to do it is to obtain an equation
analogous to eq.(\ref{InteqNphi}) and to see that it can be satisfied by
asymptotically   constant $N$ and $\phi$ only for $a=0$. This is just the
Einstein-Higgs system with a  massless real scalar.

\section{Conclusion}
We have analyzed in detail, using both numerical  and analytic methods, the
cosmic
string solutions in dilaton-gravity. Contrary to previous works, we did not  
limit ourselves  
to one branch in the weak-field approximation. We found it most natural to
work in the
Einstein frame, but all our results are trivially transformed into the
string frame.

We have shown that the solutions generally come in pairs. That is, any point
in parameter space gives rise to two solutions with completely
   different asymptotic behaviors for the dilaton and for the geometry.
Generally,
neither of the two branches describe asymptotically flat solutions. This
should be
contrasted with the  non-dilatonic case, where
  one branch always describes asymptotically flat solutions. Only for a very  
special coupling to the dilaton, which is related to certain compactified  
superstring theories, do asymptotically flat solutions appear.

Our aim in this paper was to consider cosmic strings in the simplest
dilaton-gravity models parametrized by a phenomenological coupling
constant $a$. In the special case of fundamental string theory,
which is related to our models for $a=-1$, there is an infinity of
curvature
corrections (controlled by the reciprocal string tension $\alpha'$) and an
infinity of
string loop corrections (controlled by the string coupling which is
proportional to $e^\phi)$. However, since our solutions for $
a=-1$ are
everywhere non-singular and with finite dilaton, we expect that they
will be only slightly modified in the full (yet unknown) quantum string
theory.

The  asymptotic behaviors of the solutions in the two branches seemed at first
to depend in an extremely complicated way on the parameters of the theory.
However,
a universal picture emerged, as explained in Section 3.

These results were obtained by coupling the dilaton "uniformly" to matter in
the string frame. For comparison, we re-analyzed the whole problem using a 
uniform coupling in the Einstein frame which may seem 
to be more natural. Among other things, the correct tensorial  
(Einstein) gravity solutions now come out in the limit of vanishing  
dilaton coupling. But otherwise 
the solutions in the two models are quite similar. 

  \pagebreak

\newpage

 \begin{figure}[!t]
   \begin{center}
   \includegraphics[width=7.8cm,angle=270]{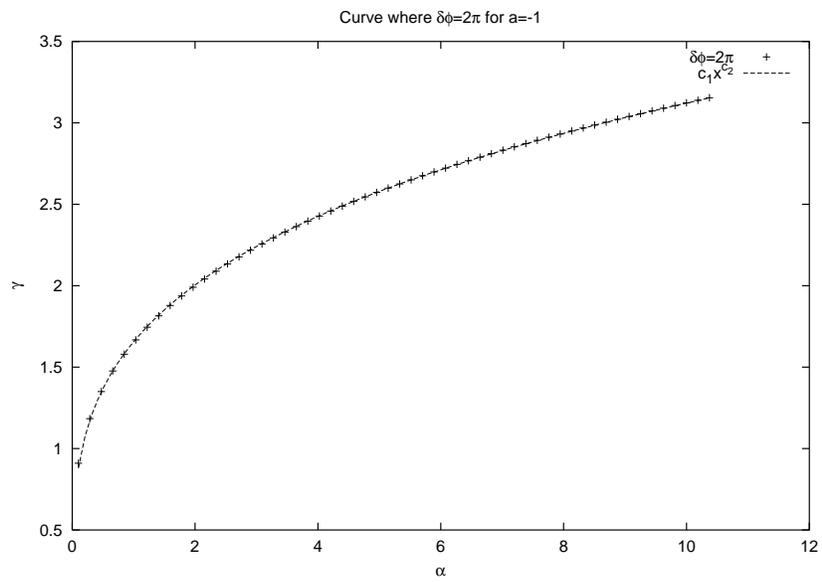}
\caption{The curve where $\delta \varphi=2\pi$ for the $a = -1$ case. The
curve is
fitted to a power-law with $c_1 \approx 1.66$ and $c_2 \approx 0.275 $.}
\label{critical}
   \end{center}
   \end{figure}

 \begin{figure}[!t]
   \begin{center}
   \includegraphics[width=7.8cm,angle=270]{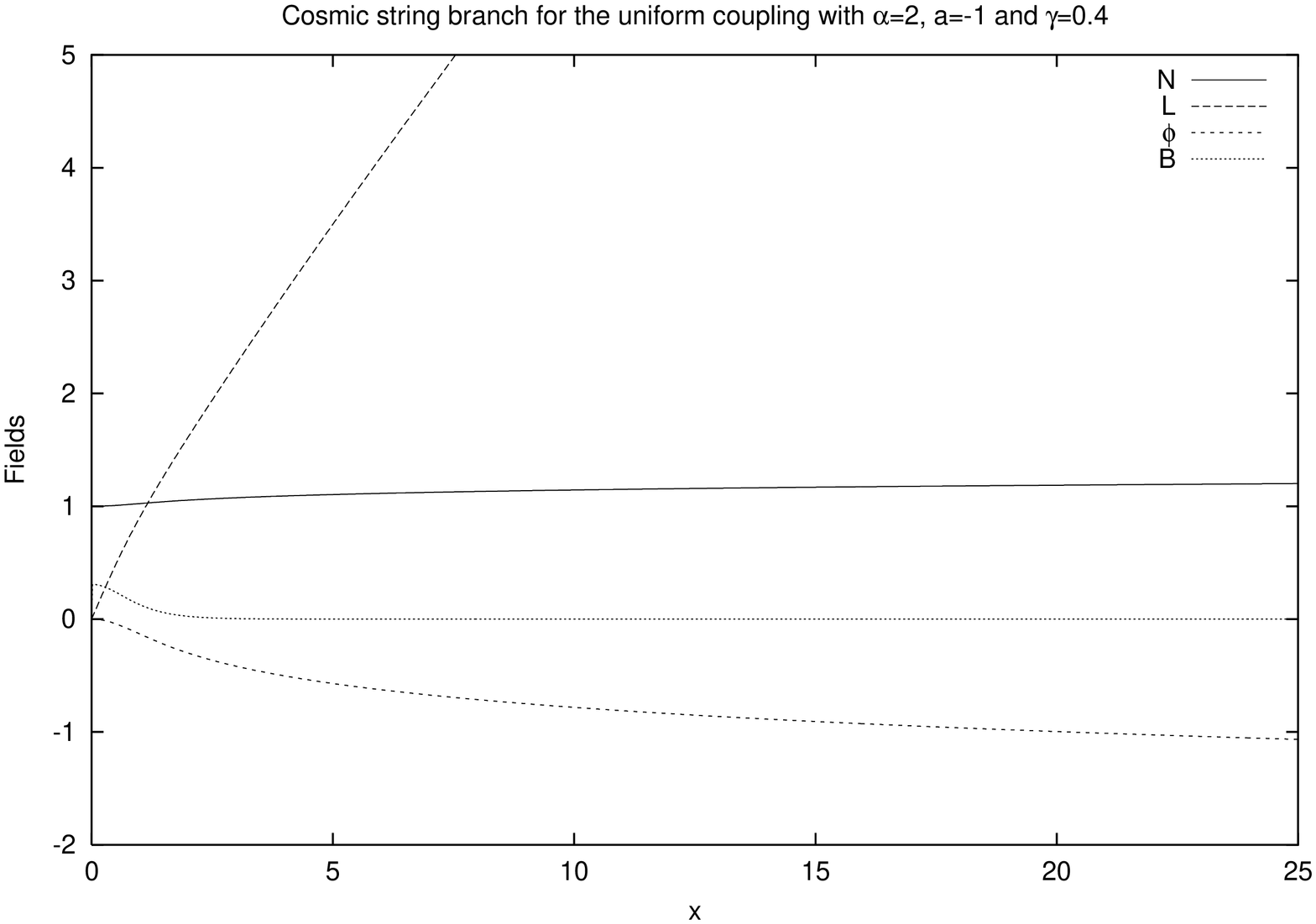}\\(a)\\
    \includegraphics[width=7.8cm,angle=270]{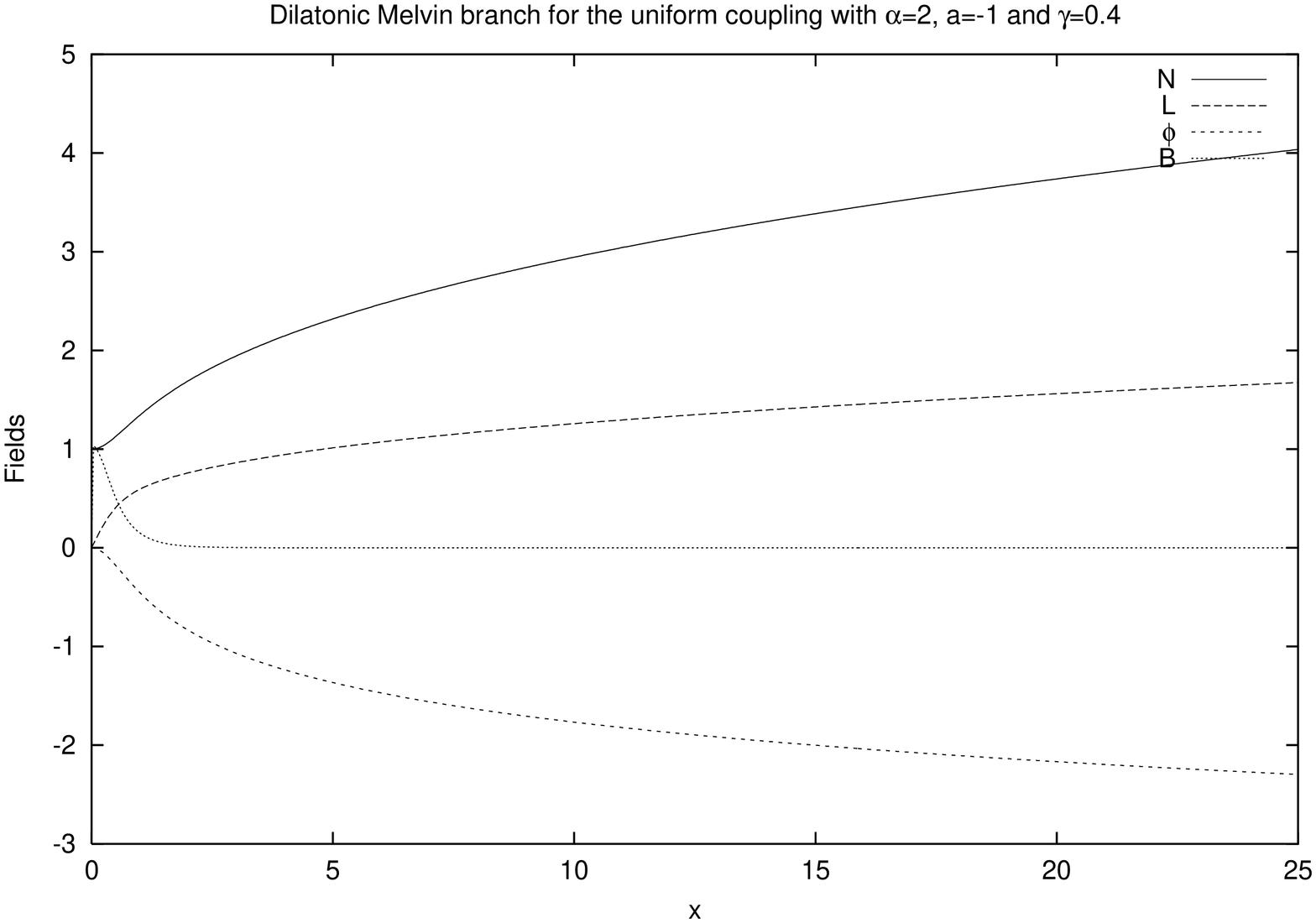}\\(b)
\caption{The solutions for uniform coupling with
 $a=-1$, $\alpha = 2$, $\gamma = 0.4$.
(a) Cosmic string branch. (b) Dilatonic Melvin branch. }
\label{SoltnUNIFa2a-1g0.4}
   \end{center}
   \end{figure}

   \end{document}